\documentclass[11pt]{article}
\usepackage{amssymb}

\usepackage{graphicx}
\usepackage{amsmath}
\usepackage{makeidx}
\usepackage{indentfirst}

\begin{document}

\title{ Super-Luminal Effects for Finsler Branes as a Way to Preserve the
Paradigm of Relativity Theories}
\date{December 26, 2012}
\author{ \textbf{Sergiu I. Vacaru}\thanks{%
sergiu.vacaru@uaic.ro,\ \ http://www.scribd.com/people/view/1455460-sergiu}
\and \textsl{\small Mathematical Physics Project IDEI, Alexandru Ioan Cuza
University, UAIC, } \\
\textsl{\small Alexandru Lapu\c sneanu street, nr. 14, Corpus R, office 323;
Ia\c si, Romania, 700057 } }

\maketitle

\begin{abstract}
Using Finsler brane solutions [see details and methods in: S. Vacaru, Class. Quant. Grav. \textbf{28} (2011) 215001], we show that  neutrinos may surpass the speed of light in vacuum which can be explained by trapping effects from gravity theories on eight dimensional (co) tangent bundles on Lorentzian manifolds to spacetimes in general and special relativity. In nonholonomic variables, the bulk gravity is described by Finsler modifications depending on velocity/ momentum coordinates. Possible super-luminal phenomena are determined by the width of locally anisotropic brane (spacetime) and induced by generating functions and integration functions and constants in coefficients of metrics and nonlinear connections. We conclude that Finsler brane gravity trapping mechanism may explain neutrino super-luminal effects and almost preserve the paradigm of Einstein relativity as the standard one for particle physics and gravity.

\vskip0.1cm

\textbf{Keywords:}\ Super--luminal travel, Finsler branes, modified gravity,
violation of local Lorentz symmetry. \vskip5pt

PACS 2008:\ 04.50.Kd, 04.90.+e, 11.25.-w, 11.30.Cp, 02.40.-k
\end{abstract}

\textbf{Introduction.} It was a very surprising result in experimental particle
physics that  neutrino may travel faster than light in vacuum \cite{opera}. Perhaps, that experiment was not correct and there are not other ones confirming it correctness. Nevertheless, that  generated in a series of  papers \cite{theoropera} with
different explanations implementing Lorentz symmetric violations, extra
dimensions, quantum corrections from different string, noncommutative,
nonlinear dispersion theories etc.  If the metrics in such theories depend on "velocity/momentum" type variables with warping/trapping, there are possible also super-luminal physical effects.

Classical and quantum physical theories with modifications of
velocity/momentum constants and variables can be naturally modelled by
modified dispersions and gravitational anisotropic polarizations of
interaction constants and fields on (co) tangent bundles. Such nonlinear
dispersion relations and related Finsler theories were studied in different
approaches oriented to applications in modern cosmology, astrophysics and
particle physics, or for elaborating generalized Finsler (super)string and
noncommutative anisotropic gravity and nonholonomic Ricci flow theories,
string/brane foam models etc \cite{ndisp,vncfinsl,lam,pfeifer}.

The goal of this letter is to show that super--luminal phenomena can be
induced by trapping effects in Finsler gravity models \cite{vfinslbr}
constructed on tangent bundles of Lorentzian manifolds for the general
relativity theory (GR) and, in particular, when the base manifold is taken
to be the Minkowski spacetime of the special relativity theory (SR). Finsler
brane solutions with trapping to standard (four dimensional, 4--d) spacetime
in GR and/or SR can be generated by diagonal and generic off--diagonal
metrics in the bulk (8--d) total space. Following this approach, we can
perform a ''canonical'' Finsler generalization of axiomatics formulated for
GR and SR which do not change the paradigm of standard relativity theories. %
\vskip5pt

\textbf{On Principles of Einstein--Finsler Gravity \cite{vrevfg,vexsol}:}
Let us denote by $\mathbf{V}$ and, in particular, $M$, respectively, a 4--d
Lorentz manifold (spacetime) and a Minkowski spacetime endowed with metrics $%
g_{ij}(x^{k})$ and $\eta _{ij}$ of signature $(-+++)$. Any compatible with
experimental data generalizations to locally anisotropic metrics $%
g_{ij}(x^{k},y^{a})$ depending on velocities $y^{a}$, or momenta $p_{a}$,%
\footnote{%
by $x=(x^{i})$ we denote local coordinates on $\mathbf{V}$ (indices $%
i,j,k... $ and $a,b,c....$ are coordinate or abstract ones which may take
values 1,2,3,4.} should be defined by "small" modifications of
geometric/physical objects from base spacetime manifold $\mathbf{V}$ to
tangent bundle $T\mathbf{V}$ (in particular, on $TM$). In total space, we
shall label the local coordinates in the form $u= (x,y)=\{u^{\alpha
}=(x^{i},y^{a})\}.$ We can introduce the so--called Finsler variables both
on $\mathbf{V}$ and $T\mathbf{V}$ when up to frame/coordinate transforms
\begin{equation}
\tilde{g}_{ij}(x^{k},y^{a})\sim \partial ^{2}F^{2}/\partial y^{i}\partial
y^{j}  \label{fvm}
\end{equation}
\ is determined by the non--degenerate Hessian of a generating (fundamental)
Finsler function $F(x,y)$. Such a function (also called a Finsler metric) is
usually chosen to be homogeneous on $y$--variables\footnote{%
various nonhomogeneous generating functions (for instance, regular
Lagrangians) and with other type homogeneity are also considered in modern
literature} and satisfies different conditions in ''standard'' Finsler
geometry with signature $(++++),$ or for pseudo--Euclidean local signatures.
For simplicity, we shall use the term Finsler for all possible signatures if
that will not result in ambiguities, see details in Refs. \cite%
{vrevfg,vexsol}.

The nonlinear quadratic element $ds^{2}=F^{2}(x,y)$ in Finsler geometry
generalizes the quadratic one in pseudo--Riemannian geometry, when $%
ds^{2}=g_{ij}(x)y^{i}y^{j}.$ In the second case, we can state $%
g_{ij}(x_{(0)}^{k})=\eta _{ij}$ in any point $x_{(0)}^{i}\in $ $\mathbf{V}$
and such a geometry is characterized by a unique torsionless and metric
compatible connection $\nabla $ (the Levi--Civita connection). A spacetime
in GR is determined by some data $(g_{ij}(x),\nabla (x))$ as a solution of
the\ Einstein gravitational field equations, $R_{ik}-\frac{1}{2}%
g_{ik}R=\varkappa T_{ik}$. In these formulas, $R_{ik}$ and $R$ are,
respectively, the Ricci tensor and scalar curvature of $\nabla $ and $T_{ik}$
is the energy--momentum tensor for matter fields. The constant $\varkappa $
is determined by the Newton constant and the speed of light $c$ is taken to
be constant in vacuum. An important property of the pseudo--Riemann geometry
is that geodesics (i.e. extremals) of $g_{ij}$ are equivalent to
auto--parallels of $\nabla .$

Prescribing a Finsler generating function $F(x,y),$ we do not define
completely a geometric and/or physical model on $T\mathbf{V,}$ see details
and discussions in \cite{vrevfg,vcrit}. Finsler spaces and related gravity
models are with more rich geometric structures than the (pseudo) Riemann/
Lorentzian geometry. We need additional assumptions, following some
geometric criteria and/or physical arguments, on classes of "admissible"
total and base/fiber frames of references, metrics and linear connections.
In brief, such considerations result in a statement that a Finsler
geometry/gravity model is defined by data $(F:\mathbf{N,g,D})$ following
certain conditions for three fundamental geometric objects (instead of one, $%
g_{ij}$, in GR):

\begin{enumerate}
\item $\mathbf{N=\{N}_{i}^{a}(u)\mathbf{\}}$ is a nonlinear connection
(N--connection) which, up--to frame/coordinate transforms, is defined by
semi--spray configurations, i.e. nonlinear geodesics which are equivalent to
the Euler--Lagrange equations for the effective regular Lagrangian $L=F^{2}.$
This states a nonholonomic (non--integrable) splitting $N:TT\mathbf{V=}hT%
\mathbf{V\oplus }vT\mathbf{V}$ into conventional horizontal (h) and vertical
(v) components, for an absolute (Whitney) sum $\mathbf{\oplus .} $ We can
work in N--adapted form on $T\mathbf{V}$ if there are used the so--called
N--elongated partial derivatives and differentials (equivalently,
distinguished, d, frames and co--frames), {\small
\begin{equation}
\mathbf{e}_{\nu } =\left( \mathbf{e}_{i}=\partial _{i}-N_{i}^{a}\partial
_{a},e_{a}=\partial _{a}\right),\ \mathbf{e}^{\mu } = \left( e^{i}=dx^{i},%
\mathbf{e}^{a}=dy^{a}+N_{i}^{a}dx^{i}\right) .  \label{ddif}
\end{equation}%
}

\item $\mathbf{g}=\{\mathbf{g}_{\alpha \beta }(u)\}$ is the metric in total
space $T\mathbf{V}$ defined by $F$ and $\mathbf{N,}$ usually via Sasaki
lifts of $\tilde{g}_{ij}$ (\ref{fvm}) and possible further frame/coordinate
transforms to respective h- an v--metric, $g_{ij}$ and $h_{ab}$, when
\begin{equation}
\ \ \mathbf{g}=\ g_{ij}(x,y)dx^{i}\otimes dx^{j}+h_{ab}(x,y)\mathbf{e}%
^{a}\otimes \mathbf{e}^{b}  \label{sm}
\end{equation}

\item $\mathbf{D}$ is a distinguished connection (d--connection), i.e.
linear connection preserving under parallelism the N--connection
h-v--splitting. Realistic physical theories with Dirac operators and
constructive--axiomatic approximations can be formulated for any metric
compatible, $\mathbf{Dg}=0\mathbf{,}$ when all values in distortion relation
$\mathbf{D=\nabla +Z,}$ with distortion tensor $\mathbf{Z}$ being a
combination of coefficients of torsion of $\mathbf{D,}$ are completely
defined by $\mathbf{g,}$ which (in its turn) is induced by $F$, see details
in \cite{vrevfg,vcrit}.
\end{enumerate}

In general, Finsler geometries/theories are with metric non--compatible
d--connections, $\mathbf{Dg}\neq 0.$ A typical example is that of Chern
(also known as Rund) d--connection. Gravity and matter field models with
metric noncompatible connection are less "relevant" to standard theories of
particle physics because such locally anisotropic models do not allow
''simple'' definitions of spinors and Finsler--Dirac operators, conservation
laws etc, see rigorous definitions and exact formulas for all geometric
objects and critical remarks in \cite{vcrit} and references therein.%
\footnote{%
We consider that readers are familiar with basic concepts and constructions
for the geometry of Lorentz manifolds and Finsler spaces; in this letter, we
can not provide, explain and motivate all results and methods.} To construct
physically realistic models of Finsler classical and quantum gravity,
locally anisotropic (non) commutative/ supersymmetric generalizations of
string/brane/ nonholonomic Ricci flow theories etc is more convenient to use
the Cartan d--connection, $\mathbf{D=\tilde{D}}$. This Finsler connection
also defines a canonical almost K\"{a}hler structure and allows us to
perform various types of deformation, A--brane and two connection
quantization \cite{vquant}. The canonical d--connection $\mathbf{D=\hat{D}}$
is more convenient if we try to integrate the gravitational field equations
in most general forms \cite{vexsol} and preserve similarity with the GR
theory. We can construct certain physically important classes of exact
solutions for $\mathbf{\hat{D}}$ and then to perform nonholonomic
deformations to $\mathbf{\tilde{D},}$ or to impose nonholonomic constraints
on $N_{i}^{a}$ when $\mathbf{\hat{D}}$ and $\nabla $ are described by the
same sets of coefficients in N--adapted frames (\ref{ddif}).

An \textit{Einstein--Finsler gravity (EFG)} model can be elaborated on $T%
\mathbf{V}$ following the same principles and axioms as for GR on $\mathbf{V}
$ but, in our approach, for the data $(F:\mathbf{N,g,\hat{D}}),$ see details
in Refs. \cite{vrevfg}. Here we note that the data $(F:\mathbf{N,g,\tilde{D}}%
)$ were used in \cite{vquant}, for quantum Finsler gravity models, and in %
\cite{vfinslbr}, for Finsler brane solutions. For purposes of this work, it
is enough to generalize the concept of Einstein manifold, when the
gravitational field equations are of type $R_{ik}=\Lambda g_{ik},$ to that
of Einstein--Finsler spaces determined by total metrics $\mathbf{g}_{\alpha
\beta }$ as solutions of equations
\begin{equation}
\mathbf{\hat{R}}_{~\beta }^{\alpha }=\mathbf{\Upsilon }_{~\beta }^{\alpha },
\label{efsp}
\end{equation}%
where the source $\mathbf{\Upsilon }_{~\beta }^{\alpha }(u)$ is generated
via gravitational polarization of $\Lambda \delta _{~\beta }^{\alpha }$ (see
below). In above formulas, the Ricci tensor $\mathbf{\hat{R}}_{~\beta
}^{\alpha }$ is constructed for $\mathbf{\hat{D}}$;\ $\Lambda $ is a
cosmological constant in the bulk $T\mathbf{V}$ and $\delta _{~\beta
}^{\alpha }$ is the Kronecker symbol.

\vskip5pt

\textbf{Diagonal Solutions with Trapping for Finsler Branes}. The simplest
example of tangent bundle solution on $TM$ which defines a trivial Finsler
brane configuration and trapping scenario with vanishing N--connection
structure is given by ansatz, $\mathbf{g} =$ {\small
\begin{equation}
\phi ^{2}(y^{5})\eta _{ij}dx^{i}\otimes dx^{j}- \left( \mathit{l}_{P}\right)
^{2}\overline{h}(y^{5})[\ dy^{5}\otimes \ dy^{5}+dy^{6}\otimes \ dy^{6}\pm
dy^{7}\otimes \ dy^{7}\pm dy^{8}\otimes \ dy^{8}],  \label{diagans8d}
\end{equation}%
} where $l_{P}$ is an effective analog Planck length (in the bulk $T\mathbf{V%
}$, one could be polarizations of physical constants, see \cite%
{vexsol,vfinslbr}), $\eta _{ij}=diag[-1,1,1,1];$ $\alpha ,\beta ,...=1,2,3,4$
and the fiber coordinates $y^{5},y^{6},y^{7},y^{8}$ are velocity/ momentum
type. This metric is a solution of EFG field equations (\ref{efsp}) for
sources
\begin{equation}
\Upsilon _{\ \delta }^{\beta }=\Lambda -\mathring{M}^{-(m+2)}\overline{K}%
_{1}(y^{5}),\ \Upsilon _{\ 5}^{5}=\Upsilon _{\ 6}^{6}=\Lambda -\mathring{M}%
^{-(m+2)}\overline{K}_{2}(y^{5}),  \label{source3}
\end{equation}%
where $\mathring{M}$ is a fundamental mass scale on $T\mathbf{V}$ and $%
m=2,3,4$\textbf{\ }defines the number of non--compactified ''extra''
dimensions of velocity type if {\small
\begin{equation}
\phi ^{2}(y^{5})=\frac{3\epsilon ^{2}+a(y^{5})^{2}}{3\epsilon
^{2}+(y^{5})^{2}}\mbox{ and }\mathit{l}_{P}\sqrt{|\overline{h}(y^{5})|}=%
\frac{9\epsilon ^{4}}{\left[ 3\epsilon ^{2}+(y^{5})^{2}\right] ^{2}}.
\label{cond1}
\end{equation}%
} In above formulas, $a$ is an integration constant and the width of brane
is $\epsilon ,$ with some fixed integration parameters when $\frac{\partial
^{2}\phi }{\partial (y^{5})^{2}}\mid _{y^{5}=\epsilon }=0$ and $\mathit{l}%
_{P}\sqrt{|\overline{h}(y^{5})|}\mid _{y^{5}=0}=1;$ this states the
conditions that on diagonal branes the Minkowski metric on $TV$ is 6--d or
8--d. The compatibility of such sources, field equations and conservation
laws in Finsler spaces are studied in \cite{vfinslbr} (see there explicit
formulas for $\overline{K}_{1}$ and $\overline{K}_{2}$). Such constructions
for $m=2$ are similar to those for usual 6--d diagonal brane solutions with
trapping and other various warping mechanisms and nonlinear wave
interactions \cite{branetrap}. Nevertheless, there are substantial
differences between Finsler branes and extra dimension branes. In our case,
the width $\epsilon ^{2}=40\mathring{M}^{4}/3\Lambda $ is for a brane in a
bulk with extra velocity/ momentum coordinates and certain constants are
related to $\mathit{l}_{P}$; the variable $y^{5}$ has a finite maximal value
$y_{0}^{5}$ on $TM$. \vskip5pt

\textbf{Off--Diagonal Solutions for Generic Finsler Branes}. We construct a
class of generic off--diagonal Finsler brane solutions for the canonical
d--connection $\mathbf{\hat{D}.}$ For such classes of locally anisotropic
spacetimes, and via corresponding trapping mechanisms, there are
possibilities to generate various classes of black hole, wormhole,
cosmological metrics etc.

Let us consider a metric $\mathbf{g}=\{\mathbf{g}_{\alpha ^{\prime
}\beta^{\prime }}\}$ which via frame transforms on $T\mathbf{V,}$ $\mathbf{g}%
_{\alpha ^{\prime }\beta ^{\prime }}=e_{~\alpha ^{\prime }}^{\alpha
}(u)e_{~\beta ^{\prime }}^{\beta }(u)\widetilde{\mathbf{g}}_{\alpha \beta },$
can be transformed into a generic off--diagonal metric $\widetilde{\mathbf{g}%
}=\{\widetilde{\mathbf{g}}_{\alpha \beta }=[g_{ij},h_{ab}]\}.$ The ansatz
for $\widetilde{\mathbf{g}}$ is taken in the form
\begin{eqnarray}
\widetilde{\mathbf{g}} &=&\phi ^{2}(y^{5})[g_{1}dx^{1}\otimes
dx^{1}+g_{2}dx^{2}\otimes dx^{2}+h_{3}\mathbf{e}^{3}{\otimes }\mathbf{e}%
^{3}\ +h_{4}\mathbf{e}^{4}{\otimes }\mathbf{e}^{4}\ ]  \label{fbr} \\
&&+\left( \mathit{l}_{P}\right) ^{2}~\overline{h}(y^{5})[h_{5}\mathbf{e}%
^{5}\otimes \ \mathbf{e}^{5}+h_{6}\mathbf{e}^{6}\otimes \ \mathbf{e}%
^{6}+h_{7}\mathbf{e}^{7}\otimes \ \mathbf{e}^{7}+h_{8}\mathbf{e}^{8}\otimes
\ \mathbf{e}^{8}],  \notag
\end{eqnarray}%
where the coordinates are parametrized in the form $u^{\alpha
}=(x^{1},x^{2},y^{3}=v,y^{4},y^{5}=\widetilde{v},y^{6},y^{7}=\underline{v}%
,y^{8}),$ the functions $\phi $ and $\overline{h}$ are those from (\ref%
{cond1}),\footnote{%
for simplicity, we shall consider solutions with trapping on velocity type
coordinate $y^{5},$ but the constructions can be performed in a similar for
for other fiber variables} and $e_{~\alpha ^{\prime }}^{\alpha }$ has
nontrivial coefficients $\ e_{~i^{\prime }}^{i}=\pm \phi ~\delta
_{~i^{\prime }}^{i}$ and $e_{~a^{\prime }}^{a}=\pm \mathit{l}_{P}\sqrt{|%
\overline{h}(y^{5})|}~\delta _{~a^{\prime }}^{a}.$ The metric $\widetilde{%
\mathbf{g}}$ is of type (\ref{sm}) which allows us to construct exact
solutions of EFG field equations (\ref{efsp}) following the anholonomic
deformation method \cite{vexsol,vfinslbr}. In a more simple form, Finsler
brane solutions can be generated for metrics with Killing symmetries on $%
\partial /y^{6}$ and $\partial /y^{8}$. In general, our method allows us to
find solutions depending on all 8 coordinates but such constructions are
very sophisticate and with less clear physical interpretation.

The coefficients of h--metric $g_{ij}=[g_{1}(x^{\widehat{k}}),g_{2}(x^{%
\widehat{k}}),h_{3}(x^{\widehat{k}},v),$ $h_{4}(x^{\widehat{k}},v)],$ $%
\widehat{k}=1,2,$ are parametrized with respect to a dual N--adapted basis
{\small
\begin{eqnarray*}
\mathbf{e}^{i} &=&(e^{1}=dx^{1},e^{2}=dx^{2},\ \mathbf{e}%
^{3}=dx^{3}+w_{1}(x^{\widehat{k}},v)dx^{1}+w_{2}(x^{\widehat{k}},v)dx^{2}, \\
&&\mathbf{e}^{4}=dx^{4}+n_{1}(x^{\widehat{k}},v)dx^{1}+n_{2}(x^{\widehat{k}%
},v)dx^{2}).
\end{eqnarray*}%
} The coefficients of v--metric $h_{ab}=[h_{5}(x^{\widehat{k}},\widetilde{v}%
),h_{6}(x^{\widehat{k}},\widetilde{v}),h_{7}(x^{\widehat{k}},\underline{v}),$
$h_{8}(x^{\widehat{k}},\underline{v})]$ are parametrized with respect to
{\small
\begin{eqnarray*}
\mathbf{e}^{a} &=&(\mathbf{e}^{5}=dy^{5}+\widetilde{w}_{\widehat{i}}(x^{%
\widehat{k}},\widetilde{v})dx^{\widehat{i}},\mathbf{e}^{6}=dy^{6}+\widetilde{%
n}_{\widehat{i}}(x^{\widehat{k}},\widetilde{v})dx^{\widehat{i}}, \\
\mathbf{e}^{7} &=&dy^{7}+\underline{w}_{\widehat{i}}(x^{\widehat{k}},%
\underline{v})dx^{\widehat{i}},~\mathbf{e}^{8}=dy^{8}+\underline{n}_{%
\widehat{i}}(x^{\widehat{k}},\underline{v})dx^{\widehat{i}}).
\end{eqnarray*}%
} The sources are any general ones which up to frame transforms can be
written in N--adapted form in a form $\mathbf{\Upsilon }_{\ \delta }^{\beta
}=$ {\small
\begin{equation}
[\mathbf{\Upsilon }_{\ 1}^{1}=\mathbf{\Upsilon }_{\ 2}^{2}=\mathbf{\Upsilon }%
_{2},\mathbf{\Upsilon }_{\ 3}^{3}=\mathbf{\Upsilon }_{\ 4}^{4}=\mathbf{%
\Upsilon }_{4},\mathbf{\Upsilon }_{\ 5}^{5}=\mathbf{\Upsilon }_{\ 6}^{6}=%
\mathbf{\Upsilon }_{6},\mathbf{\Upsilon }_{\ 7}^{7}=\mathbf{\Upsilon }_{\
8}^{8}=\mathbf{\Upsilon }_{8}],  \label{sourcb}
\end{equation}%
} with coefficients subjected to algebraic conditions when $\ ^{h}\Lambda
(x^{\widehat{k}})=\mathbf{\Upsilon }_{4}+\mathbf{\Upsilon }_{6}+\mathbf{%
\Upsilon }_{8},$ $\ ^{v}\Lambda (x^{\widehat{k}},v)=\mathbf{\Upsilon }_{2}+%
\mathbf{\Upsilon }_{6}+\mathbf{\Upsilon }_{8},$ $\ ^{5}\Lambda (x^{\widehat{k%
}},\widetilde{v})=\mathbf{\Upsilon }_{2}+\mathbf{\Upsilon }_{4}+\mathbf{%
\Upsilon }_{8},$ $\ \ ^{7}\Lambda (x^{\widehat{k}},\underline{v})=\mathbf{%
\Upsilon }_{2}+\mathbf{\Upsilon }_{4}+\mathbf{\Upsilon }_{6}.$ For zero
N---coefficients, such sources determine gravitational polarizations of
''constant'' cosmological constant and sources (\ref{source3})).

For data $(\widetilde{\mathbf{g}},\mathbf{\hat{D}})$ corresponding to ansatz
(\ref{fbr}), the nontrivial coefficients of the Ricci tensor $\widehat{R}%
_{~\beta }^{\alpha }$ for the system (\ref{efsp}) are computed \cite%
{vexsol,vfinslbr} {\small
\begin{eqnarray}
&&\widehat{R}_{1}^{1} =\widehat{R}_{2}^{2}=\frac{1}{2g_{1}g_{2}}[\frac{%
g_{1}^{\bullet }g_{2}^{\bullet }}{2g_{1}}+\frac{(g_{2}^{\bullet })^{2}}{%
2g_{2}}-g_{2}^{\bullet \bullet }+\frac{g_{1}^{^{\prime }}g_{2}^{^{\prime }}}{%
2g_{2}}+\frac{(g_{1}^{^{\prime }})^{2}}{2g_{1}}-g_{1}^{^{\prime \prime
}}]=-\ ^{h}\Lambda,  \label{ep1b} \\
&&\widehat{R}_{3}^{3} =\widehat{R}_{4}^{4}=\frac{1}{2h_{3}h_{4}}%
[-h_{4}^{\ast \ast }+\frac{\left( h_{4}^{\ast }\right) ^{2}}{2h_{4}}+\frac{%
h_{3}^{\ast }\ h_{4}^{\ast }}{2h_{3}}]=\ -\ ^{v}\Lambda,  \label{ep2b} \\
&&\widehat{R}_{5}^{5} =\widehat{R}_{6}^{6}=\frac{1}{2h_{5}h_{6}}%
[-h_{6}^{\circ \circ }+\frac{\left( h_{6}^{\circ }\right) ^{2}}{2h_{6}}+%
\frac{h_{5}^{\circ }\ h_{6}^{\circ }}{2h_{5}}]=-~^{5}\Lambda,  \label{ep3b}
\\
&&\widehat{R}_{7}^{7} =\widehat{R}_{8}^{8}=\frac{1}{2h_{7}h_{8}}%
[-h_{8}^{\diamond \diamond }+\frac{\left( h_{8}^{\diamond }\right) ^{2}}{%
2h_{8}}+\frac{h_{7}^{\diamond }h_{8}^{\diamond }}{2h_{7}}]= -\ ^{7}\Lambda,
\label{ep4b} \\
&&2h_{4}\widehat{R}_{3\widehat{k}} =-w_{\widehat{k}}[-h_{4}^{\ast \ast }+%
\frac{\left( h_{4}^{\ast }\right) ^{2}}{2h_{4}}-\frac{h_{3}^{\ast }\
h_{4}^{\ast }}{2h_{3}}]+\partial _{\widehat{k}}(h_{4}^{\ast }-\ln \sqrt{%
|h_{3}h_{4}|})=0,  \label{ep2c} \\
&&2\frac{h_{3}}{h_{4}}\widehat{R}_{4\widehat{k}} =n_{\widehat{k}}^{\ast \ast
}+(2h_{3}^{\ast }-3\frac{h_{4}^{\ast }}{h_{4}})n_{\widehat{k}}^{\ast }=0,
\label{ep2d} \\
&&2h_{6}\widehat{R}_{5\widehat{k}} =-\widetilde{w}_{\widehat{k}%
}[-h_{6}^{\circ \circ }+\frac{\left( h_{6}^{\circ }\right) ^{2}}{2h_{6}}+%
\frac{h_{5}^{\circ }\ h_{6}^{\circ }}{2h_{5}}]-\partial _{\widehat{k}%
}(h_{6}^{\circ }-\ln \sqrt{|h_{5}h_{6}|})=0,  \label{ep3c} \\
&&2\frac{h_{5}}{h_{6}}\widehat{R}_{6\widehat{k}} =\widetilde{n}_{\widehat{k}%
}^{\circ \circ }+(2h_{5}^{\circ }-3\frac{h_{6}^{\circ }}{h_{6}})\widetilde{n}%
_{\widehat{k}}^{\circ }=0,  \label{ep3d} \\
&&2h_{8}\widehat{R}_{7\widehat{k}} =-\underline{w}_{\widehat{k}%
}[-h_{8}^{\diamond \diamond }+\frac{\left( h_{8}^{\diamond }\right) ^{2}}{%
2h_{8}}+\frac{h_{7}^{\diamond }\ h_{8}^{\diamond }}{2h_{7}}]-\partial _{%
\widehat{k}}(h_{8}^{\diamond }-\ln \sqrt{|h_{7}h_{8}|})=0,  \label{ep4c} \\
&&2\frac{h_{7}}{h_{8}}\widehat{R}_{8\widehat{k}} =\underline{n}_{\widehat{k}%
}^{\diamond \diamond }+(2h_{7}^{\diamond }-3\frac{h_{8}^{\diamond }}{h_{8}})%
\underline{n}_{\widehat{k}}^{\diamond }=0,  \label{ep4d}
\end{eqnarray}%
} where $\widehat{k}=1,2$ and certain partial derivatives are denoted in a
''brief'' form, $g_{1}^{\bullet }\equiv \partial g_{1}/\partial
x^{1},g_{1}^{\prime }\equiv \partial g_{1}/\partial x^{2},h_{4}^{\ast
}\equiv \partial h_{4}/\partial v,h_{5}^{\circ }\equiv \partial
h_{5}/\partial \widetilde{v},h_{7}^{\diamond }\equiv \partial h_{7}/\partial
\underline{v}.$

The above system of equations (\ref{ep1b})--(\ref{ep4d}) reflects an
important decoupling property, with respect to N--adapted frames, of field
equations for different classes of gravitational theories.\footnote{%
in our works on GR and noncommutative, metric--affine, supersymmetric,
string, brane, generalized Finsler etc extensions \cite%
{vspinoretc,vexsol,vfinslbr,vrevfg}, we provided examples when the
fundamental equations (\ref{efsp}), split into certain sub--systems of
nonlinear partial differential equations (NPDE) which can be integrated in
very general forms.} For instance, the equation (\ref{ep1b}) is for a
diagonal metric on a 2--d subspace. The equation (\ref{ep2b}) contains only
the first and second derivative on $\ast =\partial /\partial v.$
Prescribing, for instance, $h_{4}(x^{\widehat{k}},v)$ and $\ ^{v}\Lambda (x^{%
\widehat{k}},v),$ we can almost always integrate on $dv$ and find $h_{3}(x^{%
\widehat{k}},v)$ up to certain classes of generating functions, integration
functions and constants and certain parameters. Having defined both $h_{3}$
and $h_{4},$ for ''well--defined'' conditions, we can can solve the
algebraic equations (\ref{ep2c}) for $w_{\widehat{k}}(x^{\widehat{k}},v)$
and (after two integrations on $dv)$ the second order PDE (\ref{ep2d}), in
order to find $n_{\widehat{k}}(x^{\widehat{k}},v).$ Here we note that in a
similar form we can construct solutions of systems (\ref{ep3b}), (\ref{ep3c}%
) and (\ref{ep3d}) (and, respectively, of (\ref{ep4b}), (\ref{ep4c}) and (%
\ref{ep4d})) when $\ast =\partial /\partial v$ is changed into $\circ
=\partial /\partial \widetilde{v}$ (and, respectively, $\diamond =\partial
/\partial \underline{v})$ for corresponding coefficients of (\ref{fbr}).

We show how the EFG equations can be integrated in very general forms for
ansatz (\ref{fbr}) with $h_{4}^{\ast }\neq 0,h_{6}^{\circ }\neq 0$ and $%
h_{6}^{\diamond }\neq 0.$ Introducing functions {\small
\begin{eqnarray*}
g_{\widehat{k}} &=&\epsilon _{\widehat{k}}\exp \psi (x^{\widehat{i}%
}),\epsilon _{\widehat{k}}=\pm 1,\mbox{ \ depending on signature }, \\
\phi &:=&\ln |h_{4}^{\ast }/\sqrt{|h_{3}h_{4}|}|,\gamma :=[\ln
(|h_{4}|^{3/4}/|h_{3}|)]^{\ast },\alpha _{\widehat{k}}=h_{4}^{\ast }\partial
_{\widehat{k}}\phi ,\beta =h_{4}^{\ast }\phi ^{\ast }, \\
\widetilde{\phi } &:=&\ln |h_{6}^{\circ }/\sqrt{|h_{5}h_{6}|}|,\widetilde{%
\gamma }:=[\ln (|h_{6}|^{3/4}/|h_{5}|)]^{\circ },\widetilde{\alpha }_{%
\widehat{k}}=h_{6}^{\circ }\partial _{\widehat{k}}\widetilde{\phi },%
\widetilde{\beta }=h_{6}^{\circ }\widetilde{\phi }^{\circ }, \\
\underline{\phi } &:=&\ln |h_{8}^{\diamond }/\sqrt{|h_{7}h_{8}|}|,\underline{%
\gamma }:=[\ln (|h_{8}|^{3/4}/|h_{7}|)]^{\diamond },\underline{\alpha }_{%
\widehat{k}}=h_{8}^{\diamond }\partial _{\widehat{k}}\underline{\phi },%
\underline{\beta }=h_{8}^{\diamond }\underline{\phi }^{\diamond };
\end{eqnarray*}%
} for $\phi ^{\ast }\neq 0,\widetilde{\phi }^{\circ }$ $\neq 0,\underline{%
\phi }^{\diamond }\neq 0;$ in (\ref{ep1b})--(\ref{ep4d}), we get $\epsilon
_{1}\psi ^{\bullet \bullet }+\epsilon _{2}\psi ^{\prime \prime } =2\
^{h}\Lambda$,
\begin{eqnarray*}
\phi ^{\ast }h_{4}^{\ast } &=&2h_{3}h_{4}~\ ^{v}\Lambda ,~\beta w_{\widehat{k%
}}+\alpha _{\widehat{k}}=0,~n_{\widehat{i}}^{\ast \ast }+\gamma n_{\widehat{i%
}}^{\ast }=0, \\
\widetilde{\phi }^{\circ }h_{6}^{\circ } &=&2h_{5}h_{6}~\ ^{5}\Lambda ,~%
\widetilde{\beta }\widetilde{w}_{\widehat{k}}+\widetilde{\alpha }_{\widehat{k%
}}=0,~\widetilde{n}_{\widehat{i}}^{\circ \circ }+\widetilde{\gamma }%
\widetilde{n}_{\widehat{i}}^{\circ }=0, \\
\underline{\phi }^{\diamond }h_{8}^{\diamond } &=&2h_{7}h_{8}~\ ^{7}\Lambda
,~\underline{\beta }\underline{w}_{\widehat{k}}+\underline{\alpha }_{%
\widehat{k}}=0,~\underline{n}_{\widehat{i}}^{\diamond \diamond }+\underline{%
\gamma }\underline{n}_{\widehat{i}}^{\diamond }=0.
\end{eqnarray*}%
This system of equations can be integrated ''step by step''  when, finally,
the most important coefficients are expressed{\small
\begin{eqnarray}
h_{4} &=&\pm \frac{1}{4}\int dv\left| \ ^{v}\Lambda \right| ^{-1}\left(
e^{2\phi }\right) ^{\ast },\mbox{ or }=\pm \frac{1}{4\ ^{v}\Lambda }%
e^{2[\phi -~^{0}\phi (x^{\widehat{k}})]},\mbox{ if }\ ^{v}\Lambda =const;
\label{sol1b} \\
h_{3} &=&\pm \left[ \left( \sqrt{|h_{4}|}\right) ^{\ast }\right]
^{2}e^{-2\phi }=\frac{\phi ^{\ast }}{2\left| \ ^{v}\Lambda \right| }\left(
\ln |h_{4}|\right) ^{\ast },\mbox{ or }=\pm \frac{\left( \phi ^{\ast
}\right) ^{2}}{4\ ^{v}\Lambda },\mbox{ if
}\ ^{v}\Lambda =const;  \notag \\
w_{\widehat{i}} &=&-\partial _{\widehat{i}}\phi /\phi ^{\ast },~n_{\widehat{i%
}}=~^{1}n_{\widehat{i}}(x^{\widehat{k}})+~^{2}n_{\widehat{i}}(x^{\widehat{k}%
})\int dv~h_{3}~|h_{4}|^{-3/2},  \label{sol1e}
\end{eqnarray}%
\begin{eqnarray*}
h_{6} &=&\pm \frac{1}{4}\int d\widetilde{v}\left| \ ^{5}\Lambda \right|
^{-1}\left( e^{2\widehat{\phi }}\right) ^{\ast },\mbox{ or }=\pm \frac{1}{4\
^{5}\Lambda }e^{2[\widetilde{\phi }-~^{0}\widetilde{\phi }(x^{\widehat{k}%
})]},\mbox{ if }\ ^{5}\Lambda =const; \\
h_{5} &=&\pm \left[ \left( \sqrt{|h_{6}|}\right) ^{\circ }\right] ^{2}e^{-2%
\widetilde{\phi }}=\frac{\phi ^{\circ }}{2\left| \ ^{5}\Lambda \right| }%
\left( \ln |h_{6}|\right) ^{\circ },\mbox{ or }=\pm \frac{\left( \phi
^{\circ }\right) ^{2}}{4\ ^{5}\Lambda },\mbox{ if
}\ ^{5}\Lambda =const; \\
\widetilde{w}_{\widehat{i}} &=&-\partial _{\widehat{i}}\widetilde{\phi }/%
\widetilde{\phi }^{\circ },~\widetilde{n}_{\widehat{i}}=~^{1}\widetilde{n}_{%
\widehat{i}}(x^{\widehat{k}})+~^{2}\widetilde{n}_{\widehat{i}}(x^{\widehat{k}%
})\int d\widetilde{v}~h_{5}~|h_{6}|^{-3/2},
\end{eqnarray*}%
\begin{eqnarray*}
h_{8} &=&\pm \frac{1}{4}\int d\underline{v}\left| \ ^{7}\Lambda \right|
^{-1}\left( e^{2\underline{\phi }}\right) ^{\diamond },\mbox{ or }=\pm \frac{%
1}{4\ ^{7}\Lambda }e^{2[\underline{\phi }-~^{0}\underline{\phi }(x^{\widehat{%
k}})]},\mbox{ if }\ ^{7}\Lambda =const; \\
h_{7} &=&\pm \left[ \left( \sqrt{|h_{8}|}\right) ^{\diamond }\right]
^{2}e^{-2\underline{\phi }}=\frac{\phi ^{\diamond }}{2\left| \ ^{7}\Lambda
\right| }\left( \ln |h_{8}|\right) ^{\diamond },\mbox{ or }=\pm \frac{\left(
\underline{\phi }^{\diamond }\right) ^{2}}{4\ ^{7}\Lambda },\mbox{ if
}\ ^{7}\Lambda =const; \\
\underline{w}_{\widehat{i}} &=&-\partial _{\widehat{i}}\underline{\phi }/%
\underline{\phi }^{\diamond },~\underline{n}_{\widehat{i}}=~^{1}\underline{n}%
_{\widehat{i}}(x^{\widehat{k}})+~^{2}\underline{n}_{\widehat{i}}(x^{\widehat{%
k}})\int d\underline{v}~h_{7}~|h_{8}|^{-3/2}.
\end{eqnarray*}%
} Above presented classes of generic off--diagonal solutions depend
respectively on generating functions $\psi (x^{\widehat{k}}),\phi (x^{%
\widehat{k}},v),\widetilde{\phi }(x^{\widehat{k}},\widetilde{v}),\underline{%
\phi }(x^{\widehat{k}},\underline{v})$ and on integration
functions/constants $~^{0}\phi (x^{\widehat{k}}),~^{0}\widetilde{\phi }(x^{%
\widehat{k}}),~^{0}\underline{\phi }(x^{\widehat{k}}),~^{1}n_{\widehat{i}%
}(x^{\widehat{k}}),~^{2}n_{\widehat{i}}(x^{\widehat{k}}),$ $\ ^{1}\widetilde{%
n}_{\widehat{i}}(x^{\widehat{k}}),~^{2}\widetilde{n}_{\widehat{i}}(x^{%
\widehat{k}}),\ ^{1}\underline{n}_{\widehat{i}}(x^{\widehat{k}}),~^{2}%
\underline{n}_{\widehat{i}}(x^{\widehat{k}})$ etc. Such values should be
defined in explicit form from certain physical considerations by imposing
additional symmetries (for instance, ellipsoid/torus configurations,
solitons etc), boundary conditions, Cauchy problem etc. In a more general
context, the solutions may depend on constant commutative and noncommutative
parameters, nonlinear solitonic hierarchies etc, see examples in Ref. \cite%
{vexsol}. It should be noted that the limit $\Lambda \rightarrow 0$ may be
not smooth for certain classes of solutions. As particular cases, such
metrics and generalize Finsler and other type connections describe
deformations from black hole, wormhole, cosmological metric in GR and
various modifications (we can use examples from \cite%
{vexsol,vfinslbr,vrevfg,vcrit,vspinoretc} etc).

The (generalized) Finsler configurations determined by any nontrivial (\ref%
{sol1b}), (\ref{sol1e}) etc are positively with nontrivial torsion induced
completely by the metric structure (\ref{fbr}) via corresponding nonhlonomic
effect of N--connection and anholonomic N--adapted frames. This is the
property of the canonical d--connection $\mathbf{\hat{D}}$ (and of the
Cartan d--connection $\mathbf{\tilde{D}}$). Imposing additional nonholonomic
constraints
\begin{eqnarray}
w_{\widehat{i}}^{\ast } &=&(\partial _{\widehat{i}}-w_{\widehat{i}})\ln
|h_{4}|,\partial _{\widehat{k}}w_{\widehat{i}}=\partial _{\widehat{i}}w_{%
\widehat{k}},\partial _{\widehat{i}}n_{\widehat{k}}=\partial _{\widehat{k}%
}n_{\widehat{i}},  \label{lccondm} \\
\widetilde{w}_{\widehat{i}}^{\circ } &=&(\partial _{\widehat{i}}-\widetilde{w%
}_{\widehat{i}})\ln |h_{6}|,\partial _{\widehat{k}}\widetilde{w}_{\widehat{i}%
}=\partial _{\widehat{i}}\widetilde{w}_{\widehat{k}},\partial _{\widehat{i}}%
\widetilde{n}_{\widehat{k}}=\partial _{\widehat{k}}\widetilde{n}_{\widehat{i}%
},  \notag \\
\underline{w}_{\widehat{i}}^{\diamond } &=&(\partial _{\widehat{i}}-%
\underline{w}_{\widehat{i}})\ln |h_{8}|,\partial _{\widehat{k}}\underline{w}%
_{\widehat{i}}=\partial _{\widehat{i}}\underline{w}_{\widehat{k}},\partial _{%
\widehat{i}}\underline{n}_{\widehat{k}}=\partial _{\widehat{k}}\underline{n}%
_{\widehat{i}},  \notag
\end{eqnarray}%
we can generate EFG solutions with zero torsion on $T\mathbf{V}$ which
describe models of GR extended on its tangent bundle to Lorentz manifolds.
If the so--called Levi--Civita conditions (\ref{lccondm}) are satisfied, the integration
functions $~^{2}n_{\widehat{i}}(x^{\widehat{k}}),^{2}\widetilde{n}_{\widehat{%
i}}(x^{\widehat{k}}),$ $\ ^{2}\underline{n}_{\widehat{i}}(x^{\widehat{k}})$
must be fixed zero. This allows us, in principle, to distinguish by future
experiments if super--luminal and other effects are on generic Finsler
branes (with nontrivial torsion) or on extensions on $T\mathbf{V}$ with the
Levi--Civita connection $\nabla .$ \vskip5pt

\textbf{Effective Super-Luminal Speeds and Modified Dispersions}. Let us
show explicitly that trapping from Finsler branes may result in
super--luminal effects in real physical spacetime. Light rays in SR can be
parametr\-ized as $x^{i}(\varsigma )$ with a real smooth parameter $0\leq
\varsigma \leq \varsigma _{0},$ when $ds^{2}/d\varsigma ^{2}=0.$ We can
consider a ''null'' tangent vector field $y^{i}(\varsigma
)=dx^{i}/d\varsigma ,$ with $d\tau =dt/d\varsigma .$ Under general
coordinate transforms $x^{i^{\prime }}=x^{i^{\prime }}(x^{i}),$ we have $%
\eta _{ij}\rightarrow g_{i^{\prime }j^{\prime }}(x^{k}).$ The condition $%
ds^{2}/d\varsigma ^{2}=0$ holds always for propagation of light, i.e. $%
g_{i^{\prime }j^{\prime }}y^{i^{\prime }}y^{j^{\prime }}=0.$ The Minkowski
metric $\eta _{ij}=diag[-1,+1,+1,+1]$ (for $i=1,2,3,4)$ determines the
quadratic line element $ds^{2}=\eta
_{ij}dx^{i}dx^{j}=-(dx^{1})^{2}+(dx^{2})^{2}+(dx^{3})^{2}+(dx^{4})^{2}$,
with space type, $(x^{2},x^{3},x^{4}),$ and time like, $x^{1}=ct,$
coordinates where $c$ is the constant speed of light (in vacuum). We use
indices of type $\overline{i},\overline{j},...=2,3,4,$ when $c^{2}=%
\widetilde{g}_{\overline{i}\overline{j}}(x^i)y^{\overline{i}}y^{\overline{j}%
}/\tau ^{2}$. This formula can be considered in GR if $g_{\widehat{i}%
\widehat{j}}(x^{i})$ are solutions of Einstein equations. \ In EFG, such a
formula is generalized to
\begin{equation}
\widetilde{c}^{2}(u^{\alpha })=\widetilde{g}_{\overline{i}\overline{j}%
}(u^{\alpha })y^{\overline{i}}y^{\overline{j}}/\tau ^{2},  \label{lightfinsl}
\end{equation}%
where $\widetilde{g}_{\overline{i}\overline{j}}$ is determined by the
h--component of solution of (\ref{efsp}). Effective super--luminal speeds
can be ''obtained'' from diagonal solutions (\ref{diagans8d}) with $\phi
^{2}(\tilde{v})=1+(a-1)(\tilde{v})^{2}/[3\epsilon ^{2}+(\tilde{v})^{2}],$
see formula (\ref{cond1}). Here we do not consider fiber contributions in
the total metric  of order $\left( \mathit{l}_{P}\right) ^{2}$ and $%
\epsilon ^{4}$ which are of order of Planck length and with very small
modifications for thick branes. Considering that $\widetilde{g}_{\overline{i}%
\overline{j}}(u^{\alpha })\simeq \phi ^{2}\eta _{\overline{i}\overline{j}},$
we express (\ref{lightfinsl}) in the form
\begin{equation*}
\widetilde{c}^{2}(a,\tilde{v},\epsilon )=c^{2}+\triangle =c^{2}\left( 1+%
\frac{(a-1)(\tilde{v})^{2}}{3\epsilon ^{2}+(\tilde{v})^{2}}\right)
\end{equation*}%
which mean that we get super--luminal speeds if $\triangle/c^2 >0,$ when the
integration constat $a>1.$ So, if the OPERA experimental data are correct,
they may be explained as a trapping from diagonal solutions from the bulk of
a brane configuration on $TM.$

Experiments with super--luminal effects can be applied to study possible
locally anisot\-ropic generalizations of spacetime structure to $T\mathbf{V}%
. $ For instance, introducing the h--component of a generic off--diagonal
solution (\ref{fbr}) with coefficients (\ref{sol1b}) and (\ref{sol1e}) into (%
\ref{lightfinsl}) we can compute locally anisotropic oscillations, any
possible fractional and fractal spacetime structure, quantum fluctuations,
or limits to modified (increases and/or decreased) maximal values of
effective speed of light.

The above quadratic on $y^{\overline{i}}$ expressions can be re--written for
arbitrary nonlinear ones, $\check{F}^{2}(x^{k},y^{\overline{j}}).$ This way,
we can model propagation of light in anisotropic media and/or \ construct an
(aether) spacetime geometry when the 8--d bulk is derived from a 4--d
Lorenzian manifold. For such constructions, we can use nonlinear homogeneous
quadratic elements, {\small
\begin{equation}
ds^{2}=F^{2}(x^{i},y^{j})\approx -(cdt)^{2}+g_{\overline{i}\overline{j}%
}(x^{k})y^{\overline{i}}y^{\overline{j}}[1+\frac{1}{r}\frac{q_{\overline{i}%
_{1}\overline{i}_{2}...\overline{i}_{2r}}(x^{k})y^{\overline{i}_{1}}...y^{%
\overline{i}_{2r}}}{\left( g_{\overline{i}\overline{j}}(x^{k})y^{\overline{i}%
}y^{\overline{j}}\right) ^{r}}]+O(q^{2}),  \label{fbm}
\end{equation}%
}when $F(x^{i},\beta y^{j})=\beta F(x^{i},y^{j}),$ for any $\beta >0.$ It is
possible to write such a relation in a fixed point \ $x^{k}=x_{(0)}^{k},$ \
when $g_{\overline{i}\overline{j}}(x_{0}^{k})=g_{\overline{i}\overline{j}}$
and $q_{\overline{i}_{1}\overline{i}_{2}...\overline{i}_{2r}}=q_{\overline{i}%
_{1}\overline{i}_{2}...\overline{i}_{2r}}$ $(x_{0}^{k}).$ A nonlinear
element (\ref{fbm}) defines a nonlinear \ modified dispersion relation (MDR)
between the frequency $\omega $ and the wave vector $k_{i}$ of light waves,%
\begin{equation}
\omega ^{2}=c^{2}[g_{\overline{i}\overline{j}}k^{\overline{i}}k^{\overline{j}%
}]^{2}\ (1-\frac{1}{r}\frac{q_{\overline{i}_{1}\overline{i}_{2}...\overline{i%
}_{2r}}k^{\overline{i}_{1}}...k^{\overline{i}_{2r}}}{[g_{\overline{i}%
\overline{j}}k^{\overline{i}}k^{\overline{j}}]^{2r}}).  \label{disp}
\end{equation}%
As a matter of principe, we can associate an effective nonlinear Finsler
metric to any such MDR which model violations of local Lorenz invariance
(LV) when coefficients $q_{\overline{i}_{1}\overline{i}_{2}...\overline{i}%
_{2r}}$ are computed for different quantum gravity (QG) phenomenology
models, or (in this work) for super--luminal phenomena via solutions of EFG
equations.

Let us discuss some important issues on MDR, Finsler theories and modified
theories of gravity and possible implications in explanation of possible
super--luminal and QG effects. In \cite{lam}, a theoretical study of
existing experimental data was performed. Authors' conclusion was that the
coefficients $q_{\overline{i}_{1}\overline{i}_{2}...\overline{i}_{2r}}$in (%
\ref{fbm}) and related (\ref{disp}) seem to be very small and this sounds to
be very pessimistic for detecting a respective QG phenomenology and LV.
Following such considerations, we may consider that Finsler brane
super--luminal effects are in contradiction to those results.

In \cite{vfinslbr}, we emphasized that local considerations based on MDR
with kinematic effects for a Finsler metric $F(x,y)$ are necessary and very
important to study possible LV (and, in this work, super--luminal)
propagation of particles and classical and quantum interaction. But any such
way derived conclusion is not an complete one because parametrizations (\ref%
{fbm}) are "geometric gauge" dependent. Using frame/coordinate transforms
and nonholonomic deformations, $\left( F:~\mathbf{g,N,D}\right) \rightarrow
(~^{0}F:~\mathbf{\check{g},\check{N},\check{D})}$, when $\ ^{0}F$ is a
typical quadratic form in GR, possible LV and super--luminal effects are
removed into data $\left( \mathbf{\check{N},\check{D}}\right) $ modeling
nonlinear generic off--diagonal quantum, and quasi--classical, interactions
in QG.

In the bulk 8--d spacetime, the solutions of EFG are with locally
anisotro\-pic gravitational interactions and broken spacetime symmetries but
on the base spacetime manifolds, with respect to N--adapted frame of
references, the MDR (\ref{disp}) may not reflect such properties. We
constructed in explicit form such classes of solutions of noncommutative
Finsler black holes \cite{vncfinsl}, see also references therein. Physical
effects for black hole/ellipsod, Taub NUT and wormhole solutions and various
gravitational--gauge--fermion interactions (in GR, extra dimension gravity,
and EFG, see \cite{vspinoretc}) can not be studied experimentally only via
Mikelson--Morley and related MDR determined  by $F$ without further
assumptions on Finsler connections and fundamental field equations. Exact
solutions with generic off--diagonal metrics and anholonomic frames
(nonlinear connection) and induced torsion effects effects are of crucial
importance in Finsler theories and velocity/momentum variables.

Different classes of solutions depend on the type of Finsler connections,
symmetry configurations, boundary conditions etc. Nevertheless, it is
possible to conclude following general arguments and main properties of
locally anisotropic theories if possible extensions of GR on $T\mathbf{V}$
are of Finsler type, or of Kaluza--Klein (with certain compactifications of
velocity/momentum coordinates), or a variant of ''velocity'' extra dimension
Einstein theory (with zero torsion). For generic Finsler configurations, the
solutions for $n_{\widehat{i}}$ coefficients depend on velocity. If the
tangent bundle spacetime theory is of another type, it will be not such a
dependence, or it will be not detected for some values higher then the limit
of ''compactification'' on $v$--coordinates. Sure, super--luminal effects
may exist in vicinity of black holes, for cosmological metrics with locally
anisotropic generalizations etc. Physically important generic off--diagonal
metrics of type (\ref{fbr}), and more general non--Killing configurations,
should contain in certain corresponding limits black hole, homogeneous and
nonhomogeneous cosmological solutions or other metrics. Such solutions can
constructed following methods elaborated in Refs. \cite%
{vexsol,vfinslbr,vrevfg,vcrit,vspinoretc}.

\vskip5pt

\textbf{Concluding }R\textbf{emarks and Further Perspectives}. This paper was
inspired by results about possible superluminal propagation of neutrinos
(additionally to the recent OPERA result \cite{opera}, there were reported
such data by FERMILAB07--MINO, FERMILAB79 collaborations \cite{preopera}). At present, it is considered that those experiments were performed not correctly. Nevertheless, we prove theoretically that  superluminality effects may exist in a total Finsler like bulk even the  principles of relativity are correct and our physical spacetime is a Lorentz base manifold.

In our approach, we do not suggest violation of the Poincar\'{e} invariance
at the fundamental level even we addressed to Finsler spacetime geometry
models, which are naturally related to modified dispersion relations (MDR)
and violation of local Lorentz invariance (LV). Such ''radical''
possibilities with changing paradigms of spectial relativity (SR) and
general relativity (GR), violation of metric compatibility etc, were
considered in Refs. \cite{bogosl}, see also a series of articles \cite{ndisp}%
. The goal of this paper is to ask and provide an answer to the question:
what minimal changes to the standard models of physics could explain
superluminal phenomena and how this can be accommodated to existing
principles and axiomatics of Einstein relativity theories?

Any constructions related to modifications for a maximal speed of
interactions which do not change drastically the existing paradigms should
be performed for (co) tangent bundles to Lorentz and Minkowski manifolds
(respectively, in GR and SR) and this should be related to certain small
MDR. Logically, we suppose that fundamental principles are still valid on
Einstein spacetimes as brane configurations imbedded into (co) tangent
bundle bulk generalizations, with finite effective velocities and
polarizations of constants and field interactions. Such models can be always
described effectively in so--called Finsler variables. Nonholonomic
variables of Finsler and other type equivalent ones can be introduced even
in GR for non--integrable 2+2 decompositions, see details in \cite%
{vrevfg,vexsol,vcrit}. This allows us to elaborate the concept of
Einstein--Finsler gravity (EFG) gravity theory as an extension of GR to (co)
tangent Lorentz manifolds with some well defined metric compatible Finsler
connections (for instance, the Cartan and/or the canonical distinguished
ones). For this class of theories, we do not need to introduce new
definitions of Finsler space like in \cite{pfeifer}; our Finsler
constructions are dynamically derived on tangent spaces to Lorentz manifolds.

An important property of EFG field equations is that they can be integrated
in very general diagonal and generic off--diagonal forms. Using certain
classes of Finsler brane solutions constructed in \cite{vfinslbr} (in this
letter, we work with the canonical distinguished connection instead of the
Cartan and/or normal ones), we proved in explicit form that super--luminal
effects for neutrinos can be explained by trapping from the Finsler
spacetimes in the bulk to to real Minkowski and Einstein spacetimes. For
such constructions, we still keep the standard paradigm of GR and SR but
with some small modifications to include well--defined Finsler like
connections for the bulk (co) tangent Lorentz bundles.

The challenge to reconcile theoretical results on dark energy and dark matter a with absence of superluminal
propagation for supernova neutrinos still exists. To solve this problem
following our approach is necessary to consider exact solutions in EFG
theories for locally anisotropic and generic off--diagonal gravitational and
Yang--Mills--Higgs--Diract fields interactions. Such solutions seem to be
similar to those constructed in extra dimension, metric--affine and/or
noncommutative theories \cite{vspinoretc}.

Finally, we note that EFG theories can be quantized using deformation,
A--brane and two connection formalism \cite{vquant}. This provides us with
new methods which allow us to study possible quantum gravitational and
matter fields superluminal effects.

\vskip5pt

\textbf{Acknowledgments}. The research for this paper is partially supported
by the Program IDEI, PN-II-ID-PCE-2011-3-0256. I'm grateful for former
collaboration and/or important discussions/correspondence relevant to this
paper to P. Stavrinos, N. Mavromatos, C. L\"{a}mmerzahl, V. Perlick, S. Odintsov,  and C. Castro.

\end{document}